\documentclass[prl,twocolumn,amsmath,amssymb,floatfix,superscriptaddress]{revtex4-1}
\usepackage{graphicx} 
\usepackage{booktabs}
\usepackage{color}
\usepackage[version=4]{mhchem}
\usepackage{xspace, bm, braket}
\usepackage{siunitx}
\usepackage{verbatim}
\usepackage{subfigure}
\usepackage[colorlinks, citecolor=blue, linkcolor=blue, urlcolor=blue, breaklinks]{hyperref}

\newcommand{\SRO}{\ce{Sr2RuO4}}
\newcommand{\SMO}{\ce{Sr2MoO4}}
\newcommand{\etal}{\textit{et al.}}

\newcommand{\imag}{\text{i}}
\newcommand{\angstrom}{\textup{\AA}}

\def\beq{\begin{equation}}
\def\eeq{\end{equation}}

\begin{document}
	
\title{Sr$_2$MoO$_4$ and Sr$_2$RuO$_4$: Disentangling the Roles of Hund's and van~Hove Physics}

\author{Jonathan Karp}
\email{jk3986@columbia.edu}
\affiliation{Department of Applied Physics and Applied Math, Columbia University, New York, NY 10027, USA}
\author{Max Bramberger}
\affiliation{Arnold Sommerfeld Center of Theoretical Physics, Department of Physics, University of Munich, Theresienstrasse 37, 80333 Munich, Germany}
\affiliation{Munich Center for Quantum Science and Technology (MCQST), Schellingstrasse 4, 80799 Munich, Germany}
\author{Martin Grundner}
\affiliation{Arnold Sommerfeld Center of Theoretical Physics, Department of Physics, University of Munich, Theresienstrasse 37, 80333 Munich, Germany}
\affiliation{Munich Center for Quantum Science and Technology (MCQST), Schellingstrasse 4, 80799 Munich, Germany}
\author{Ulrich Schollw\"ock}
\affiliation{Arnold Sommerfeld Center of Theoretical Physics, Department of Physics, University of Munich, Theresienstrasse 37, 80333 Munich, Germany}
\affiliation{Munich Center for Quantum Science and Technology (MCQST), Schellingstrasse 4, 80799 Munich, Germany}
\author{Andrew J. Millis}
\affiliation{Department of Physics, Columbia University, New York, NY 10027, USA}
\affiliation{Center for Computational Quantum Physics, Flatiron Institute, 162 5th Avenue, New York, NY 10010, USA}
\author{Manuel Zingl}
\affiliation{Center for Computational Quantum Physics, Flatiron Institute, 162 5th Avenue, New York, NY 10010, USA}

\date{\today}

\begin{abstract}
	\SMO{} is isostructural to the unconventional superconductor \SRO{} but with two electrons instead of two holes in the Mo/Ru-t$_{2g}$ orbitals. Both materials are Hund's metals, but while \SRO{} has a van~Hove singularity in close proximity to the Fermi surface, the van~Hove singularity of \SMO{} is far from the Fermi surface. By using density functional plus dynamical mean-field theory,
	we determine the relative influence of van~Hove and Hund's metal physics on the correlation properties. We show that theoretically predicted signatures of Hund's metal physics occur on the occupied side of the electronic spectrum of \SMO{}, identifying \SMO{} as an ideal candidate system for a direct experimental confirmation of the theoretical concept of Hund's metals via photoemission spectroscopy. 
\end{abstract}

\maketitle
\SRO{} has emerged as an exemplary quantum material, providing fundamental insight into the effect of electronic correlations on material properties~\cite{Mackenzie1996b,Maeno1997,Bergemann2003,Stricker2014, Behrmann2012, Tamai2018, Deng2016, Veenstra2013, Zingl2019,Sarvestani2018,Lee2020,Zhang2016,Kim2018,Strand2019}. The rich electronic properties of \SRO{} are determined by a sophisticated interplay of factors, including the Coulomb repulsion, spin-orbit coupling, and a van~Hove singularity, but it is believed that the nontrivial physics of the interorbital Hund's interaction~\cite{Yin2011,medici11,Georges2013,Lanata2013,fanfarillo15,Stadler2015,Stadler2019,Horvat2019,Deng2019} is at the heart of the strongly correlated nature of this material~\cite{Mravlje2011,Deng2019,Lee2020,Kugler2020}. However, unambiguous experimental observation of Hund's-related physics has been challenging. For example, the presence of a van~Hove singularity in the vicinity of the Fermi level impacts electronic correlations, masking the effects of the Hund's coupling on the quasiparticle mass enhancement~\cite{Mravlje2011,Lee2020,Kugler2020}. While Hund's physics has been predicted to produce a characteristic peak in the single-particle spectrum~\cite{Stadler2019, Wadati2014, Horvat2019}, for \SRO{} this peak occurs on the unoccupied side of the spectrum~\cite{Stricker2014,Kim2018,Sarvestani2018}. Thus, a direct experimental observation with conventional photoemission spectroscopy is challenging, though indirect hints have been seen in optical conductivity~\cite{Stricker2014}.

In this Letter, we use a combination of density functional theory (DFT) and dynamical mean-field theory (DMFT) to argue that \SMO{}, a material isostructural to \SRO{} but with a different electron count, provides an ideal platform to study Hund's physics, and obtain insight into the role of van~Hove singularities. We show that (i) the characteristic Hund's metal peak appears on the occupied side of the electronic spectrum for \SMO{}, making it directly observable in conventional photoemission experiment and (ii) in contrast to \SRO{}, for \SMO{} the van~Hove singularity is substantially displaced from the Fermi surface, permitting the effects of van~Hove and Hund's physics to be disentangled.

\SMO{} crystallizes in the same tetragonal $I4/mmm$ crystal structure as \SRO{}, with $a=b$ and $c$ lattice parameters being slightly larger, as expected from the larger ionic radius of Mo$^{4+}$ in comparison to Ru$^{4+}$~\cite{shirakawa2001synthesis, Shirakawa2001improved, Ikeda2000}. The octahedral oxygen environment surrounding the Ru/Mo atoms leads to an e$_g$-t$_{2g}$ splitting of the Ru/Mo-4$d$ shell with unoccupied e$_g$ orbitals and three t$_{2g}$ orbitals occupied by 2 electrons in \SMO{} and 4 electrons in \SRO{}. Two decades ago, \SMO{} was synthesized in polycrystalline form~\cite{shirakawa2001synthesis, Shirakawa2001improved, Ikeda2000}, and later, \SI{60}{uc} single-crystal films were reported~\cite{radetinac2011single}. In contrast to the vast literature on \SRO{}, only the basic electronic structure of \SMO{} has been studied with DFT~\cite{hase2003electronic}.

\begin{figure}[t]
	\centering
	\includegraphics[width = \columnwidth]{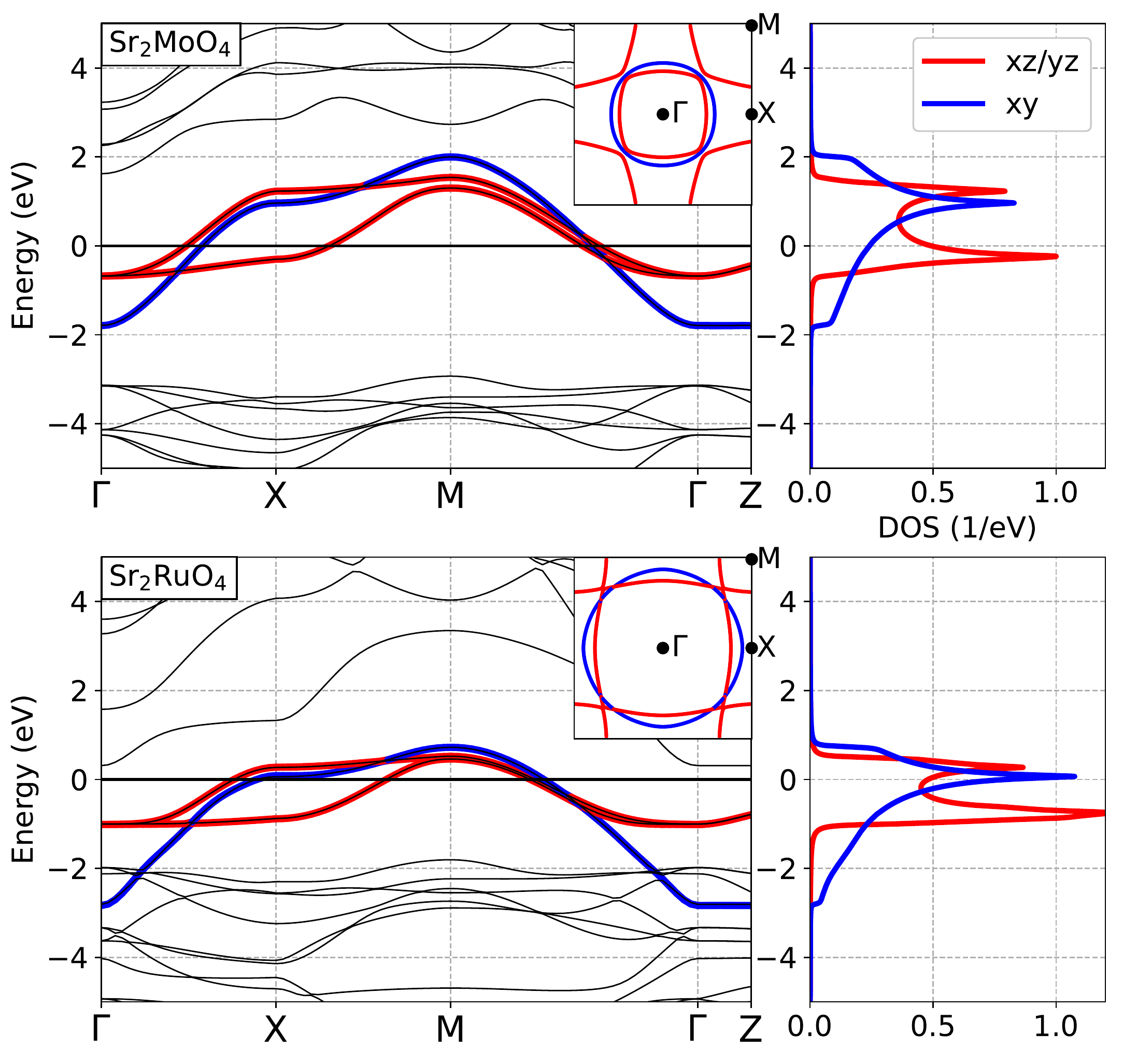}
	\caption{Left: comparison of DFT (black) and $xy$ (blue) and $xz/yz$-derived (red)  Wannier bands for \SMO{} (top) and \SRO{} (bottom). Insets: Fermi surfaces in the $k_z = 0$ plane. Right: orbitally resolved Wannier densities of states (per spin).
	}
	\label{fig:dft_bands}
\end{figure}

Fig.~\ref{fig:dft_bands} shows the DFT electronic structure calculated with Wien2k~\cite{Blaha2018} using the PBE-GGA~\cite{PBE} exchange-correlation functional and experimental atomic positions~\cite{hase2003electronic,Supp}, along with Wannier bands discussed below. The insets of Fig.~\ref{fig:dft_bands} show the Fermi surfaces, which consist of three sheets: two electron-like sheets centered at $\Gamma$ and one hole-like pocket centered at the M point. The electron sheets are smaller and the hole pockets are larger in \SMO{} than in \SRO{}, due to the lower electron count of \SMO{}. Without spin-orbit coupling, the smaller electron sheet and the hole-pockets are of pure $xz/yz$ character (red), whereas the larger electron sheet is of $xy$ orbital character (blue). 

The inclusion of spin-orbit coupling, which is slightly smaller in \SMO{} (\SI{80}{meV}) than in \SRO{} (\SI{100}{meV}), leads to a momentum-dependent mixing of the orbital character of the Fermi surface sheets~\cite{Haverkort2008, Veenstra2013,Tamai2018}. In contrast to \SRO{}, the spin-orbit coupling does not cause a restructuring of the Fermi surface in \SMO{}. We discuss the electronic structure with spin-orbit coupling in the Supplemental Material~\cite{Supp}, but we neglect it for most of this work as it is not important for the Hund's-related electronic correlations of interest here.

To capture the low energy physics, we construct a basis of three t$_{2g}$-like maximally localized Wannier orbitals~\cite{MLWF1,MLWF2,wien2wannier,wannier90}. As shown in the left-hand panels of Fig.~\ref{fig:dft_bands}, the Wannier states (colored) reproduce the DFT bands (black) very precisely in both materials. For \SMO{} the t$_{2g}$-derived bands around the Fermi energy are separated from the O-$p$ states by more than \SI{1}{eV}, which makes the selection of a low-energy subspace even more natural. The shape of the Wannier orbital density of states (DOS), Fig.~\ref{fig:dft_bands} right-hand panels, is a result of the quasi-2D crystal structure, which makes the rather 2D-like $xy$ orbital (blue) different from the more 1D $xz/yz$ ones (red). For \SMO{} the degenerate $xz/yz$ orbitals have a wider bandwidth (\SI{2.2}{eV}) than for \SRO{} (\SI{1.5}{eV}), but the difference in bandwidths of the $xy$ orbital is less (\SI{3.6}{eV} versus \SI{3.8}{eV}). Overall, the band structures and DOS of the two materials are very similar apart from a shift in the Fermi level due to the different electron count. For the $xz/yz$ orbitals this results in the upper band-edge singularity being closer to the Fermi level for \SRO{} and the lower one being closer to the Fermi level for \SMO{}. There is another important qualitative difference: For \SMO{} the saddle point of the $xy$-derived band at the X point, corresponding to a van~Hove singularity in the DOS, is at $\sim$\SI{1}{eV} above the Fermi energy, while for \SRO{} it is in close proximity to the Fermi energy. We will see in the following how this key difference in the electronic structure impacts the strength of electronic correlations. 

\begin{figure}[t]
	\centering
	\includegraphics[width = \linewidth]{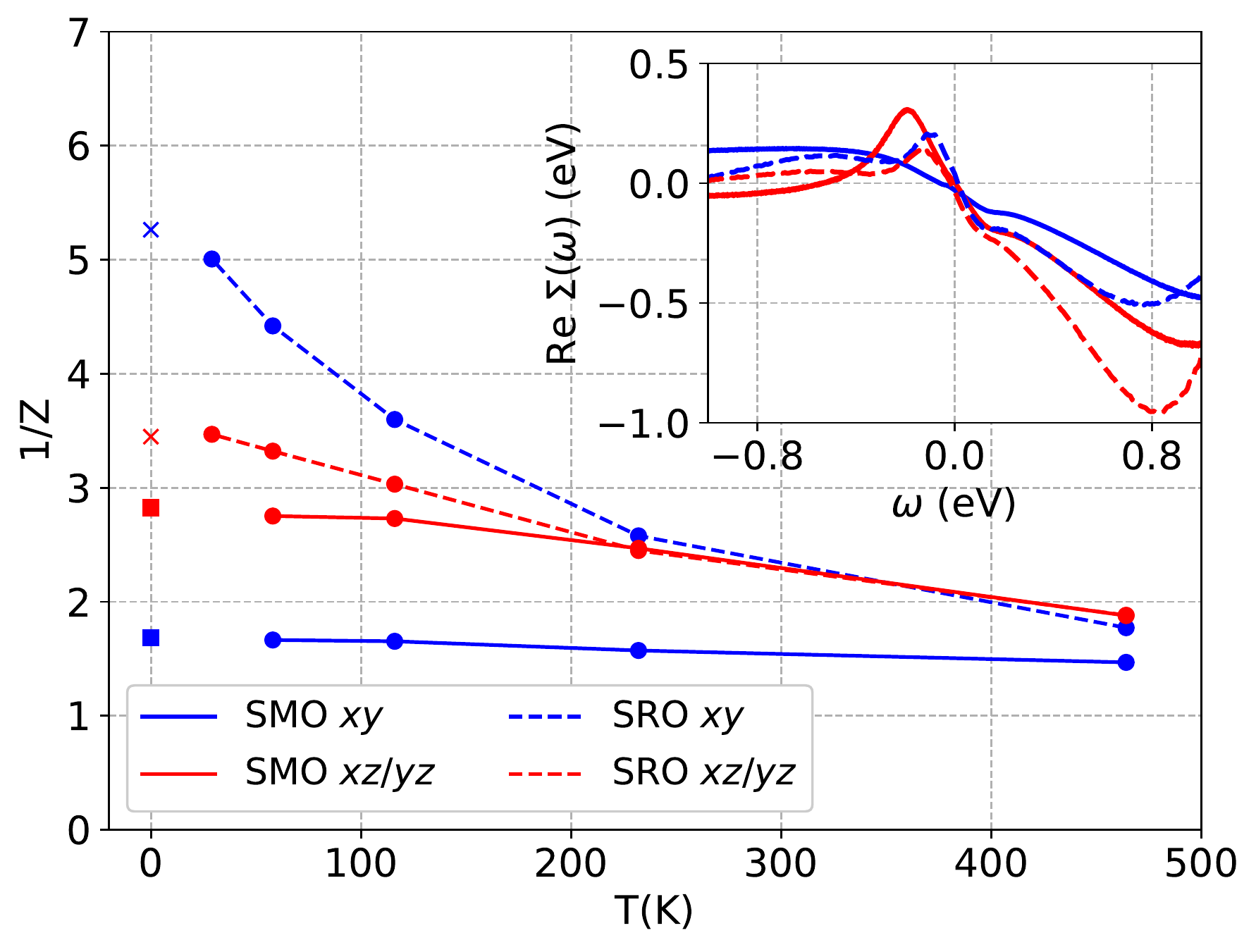}wi
	\caption{Main: DMFT mass enhancement parameters $1/Z$~\cite{Supp} for the $xy$ (blue) and $xz/yz$ (red) orbitals of \SMO{} (solid lines) and \SRO{} (dashed lines) as a function of temperature. The error bars of the CTHYB results (circles) are smaller than the marker size. The zero temperature values (squares and crosses) have been obtained using a matrix product states impurity solver~\cite{Linden2020,wolf15iii, Supp}. Inset: real part of the DMFT real-frequency self-energies obtained at $T = \SI{232}{K}$ using CTHYB as the impurity solver and with subsequent analytic continuation to the real-frequency axis~\cite{TRIQS,TRIQS/CTHYB,beach04,Supp}. Note that the chemical potential has also been subtracted. For \SRO{} (dashed lines) we show the negative of the reflection of the self-energies through $\omega=0$; i.e., $-\Sigma(-\omega)$.
	}
	\label{fig:sigma}
\end{figure}

We add local interactions of Hubbard-Kanamori form~\cite{Kanamori1963} using a Coulomb repulsion $U=\SI{2.3}{eV}$ and a Hund's coupling $J=\SI{0.4}{eV}$~\cite{fn3} and solving the resulting problem within single-site DMFT~\cite{georges96,TRIQS,TRIQS/DFTTOOLS}. We obtain results at nonzero temperatures ranging from \num{29} to \SI{464}{K} by employing the continuous-time quantum Monte Carlo method in the hybridization expansion (CTHYB)~\cite{Gull:2011lr, TRIQS/CTHYB} as the impurity solver and at effectively zero temperature using a matrix product states (MPS) based solver~\cite{wolf15iii,Linden2020}.

We characterize the strength of electronic correlations by the inverse quasiparticle renormalization $Z^{-1}=1-\partial \text{Re}\Sigma(\omega\rightarrow 0)/\partial \omega$~\cite{Supp} related, in the single-site DMFT approximation, to the quasiparticle mass enhancement as $m^\star/m=Z^{-1}$, shown in Fig.~\ref{fig:sigma}.
For both materials the calculated low-temperature mass enhancements agree with experimental specific heat measurements, which indicate that the overall mass enhancement of \SRO{} is about 4~\cite{mackenzie03, Bergemann2003, Tamai2018}, while for \SMO{} correlations are weaker and result in a mass enhancement of only around 2~\cite{Ikeda2000,hase2003electronic}. From the specific heat $c_p \sim \sum_l (m^*/m)_l \ N_l(E_F) $,  where $l\in\{xy, xz,yz\}$ and $N_l(E_F)$ is the bare DOS at the Fermi energy, we obtain a specific heat ratio $c_p^{\text{SRO}}/c_p^{\text{SMO}} = 2.4$, which is in good agreement with the experimental value of about 2.8~\cite{Ikeda2000}.

\begin{figure}[t]
	\centering
	\includegraphics[width = \columnwidth]{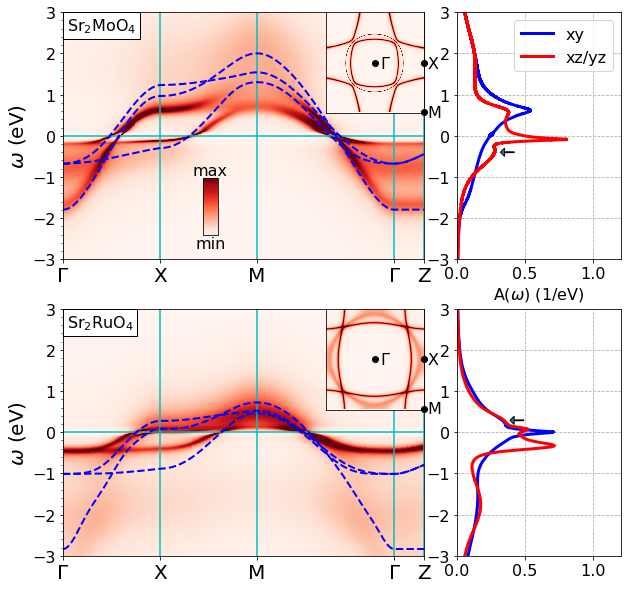}
	\caption{Many-body electronic structure obtained with DMFT for \SMO{} (top panels) and \SRO{} (bottom panels) at $T=\SI{232}{K}$. Left: momentum-resolved spectral function $A_k(\omega)$ (false color) along a high-symmetry $k$ path through the Brillouin zone compared to the Wannier bands (dashed blue lines). Insets: spectral function $A_k(\omega = 0)$ in the $k_z = 0$ plane. Right: momentum-integrated spectral function $A(\omega)$ (per spin) for the $xy$ (blue) and the $xz/yz$ (red) orbitals. Black arrows point to the Hund's peaks. Note the different range of energy in comparison to Fig.~\ref{fig:dft_bands}.}
	\label{fig:correlated_structure}
\end{figure}

At zero temperature, the $xz/yz$ orbital mass enhancements of the two materials are approximately in the same ratio as the inverses of the respective bandwidths. We attribute this finding to the nearly symmetrical shape of the $xz/yz$ DOS in both materials; see Fig.~\ref{fig:dft_bands}. The situation for the $xy$ orbital is different: For \SRO{}, in agreement with previous works~\cite{Mravlje2011,Kugler2020}, we find that even though the $xy$ orbital has the larger bandwidth, its mass enhancement is nearly twice as large as the mass enhancement of the $xz/yz$ orbitals. The unusually large $xy$ orbital mass enhancement of \SRO{} has been attributed to the proximity of the van~Hove singularity to the chemical potential~\cite{Mravlje2011,Kugler2020}. Conversely, for \SMO{} the van~Hove singularity is far removed from the chemical potential, and the mass enhancements are consistent with the difference in the bare bandwidths; the $xy$ orbital is substantially less correlated than the $xz/yz$ orbitals. As \SRO{} is cooled, the mass enhancements exhibit a strong temperature and orbital dependence with no sign of saturation above \SI{30}{K}. This is in accordance with a Fermi liquid temperature of about \SI{25}{K}~\cite{mackenzie03,Kugler2020}. For \SMO{}, we observe only a weak temperature dependence of the mass enhancement, and its saturation at about \SI{100}{K} indicates a much higher Fermi liquid coherence scale than in \SRO{}. These findings suggest that the van~Hove singularity provokes a suppression of the Fermi liquid temperature in \SRO{} and demonstrate the importance of capturing the interplay of correlation physics and specifics of band structure to understand the quasiparticle properties in strongly correlated materials.

In contrast to the van~Hove singularity, the spin-orbit coupling does not influence the mass enhancements of \SRO{}~\cite{Kim2018,Linden2020}. However, it is known from theory and experiment that electronic correlations lead to an effective spin-orbit coupling 2 times larger than its bare value~\cite{liu_prl_2008,Zhang2016,Kim2018,Tamai2018,Linden2020}. By using the MPS-based impurity solver for calculations with spin-orbit coupling, we find that the same picture holds in \SMO{}, yielding a slightly higher enhancement factor of about 2.5 (see Supplemental Material~\cite{Supp}). We therefore conclude that the correlation-enhanced spin-orbit coupling in both materials is to a large degree a result of local interactions~\cite{liu_prl_2008} rather than a consequence of van~Hove physics.

The materials' similarities and differences are also evident in the correlated spectral function, shown in the left-hand panel of Fig.~\ref{fig:correlated_structure}. We see that for \SRO{}, the unoccupied states conform closely to the bare bands, while the occupied bands are shifted substantially toward the chemical potential. For \SMO{}, the renormalization is less severe, and the unoccupied states differ considerably from the DFT bands.
In the insets of Fig.~\ref{fig:correlated_structure}, we show the spectral function at $T=\SI{232}{K}$ and $\omega = 0$ in the $k_z = 0$ plane. These many-body Fermi surfaces portray the major differences found in the $xy$ orbitals. While the $xy$ sheet is very sharp in \SMO{}, we find it to be broadened in \SRO{}. This is caused by the van~Hove singularity in \SRO{}, which is shifted even closer to the chemical potential due to electronic correlations.

Results for the orbitally resolved self-energies at $T=\SI{232}{K}$ are presented in the inset of Fig.~\ref{fig:sigma}. Note that for \SRO{} what is shown is the negative of the reflection of the self-energy through $\omega=0$, i.e. $-\Sigma(-\omega)$. The $xz/yz$ self-energies for the two materials have a clear qualitative similarity, showing that for these orbitals \SMO{} is -- to a good approximation -- indeed the particle-hole dual of \SRO{}. The self-energies have a negative slope at $\omega=0$, corresponding to the usual low-energy reduction of the quasiparticle velocity due to strong correlations. There is, however, an interesting inversion of slope around $\omega=\SI{-0.2}{eV}$, which has been pointed out in several DMFT works on \SRO{}~\cite{Stricker2014,Mravlje2016,Kim2018,Kugler2020}. For \SMO{} the inversion of slope is only present in the $xz/yz$ self-energy.

The inversion of slope occurs still well within the bare bandwidth, and may lead to a `retracted' renormalization of the quasiparticle dispersion. The consequence is an additional side peak in the spectral function $A(\omega)$ (marked with small arrows in Fig.~\ref{fig:correlated_structure}, right-hand panels), which cannot be related to a structure present in the noninteracting DOS. Model system calculations indicate that the inverted slope and the corresponding side peak in $A(\omega)$ are characteristic signatures of the spin-orbital separation occurring in Hund's metals~\cite{Stadler2015,Stadler2019,Horvat2019}. For \SMO{}, with two electrons in three orbitals, the screening of the orbital degrees of freedom requires binding a conduction band electron to the correlated site, resulting in the formation of a large $S=3/2$ local moment~\cite{Stadler2019}. Breaking this composite spin requires the removal of an electron, and thus an excitation corresponding to the energy of this process can be expected in the electron-removal part of the spectrum. Conversely, for \SRO{} with a more than half-filled shell, i.e. four electrons in three orbitals, the screening involves an additional hole, and thus the Hund's metal side peak is found at positive energies.

To our knowledge, no photoemission experiment has yet observed this side peak, probably because most studied Hund's metals have more than half-filled correlated shells so the Hund's peak is on the unoccupied side of the spectrum and not observable in photoemission. Crucially, for \SMO{} the Hund's metal peak is present on the occupied side and therefore observable in photoemission. However, in the momentum-integrated spectral function $A(\omega)$, the Hund's metal peak is a relatively weak feature. We show here how the momentum dependence of the spectral function reveals the importance of Hund's physics more clearly.

Fig.~\ref{fig:correlated_structure} shows that for \SMO{} along the $\Gamma$-X path, there are two pronounced spectral features on the occupied side, one at \SI{-0.2}{eV} corresponding to the renormalized $xz/yz$-derived bands and another corresponding to the strongly dispersing $xy$-derived band. Between these two is additional spectral weight which corresponds to the Hund's metal excitation (see also Supplemental Material~\cite{Supp}). The Hund's metal spectral weight roughly follows the energy of the lower noninteracting $xz/yz$-derived band. We also see that the occupied side of $A_k(\omega)$ of \SMO{} is very different from that of \SRO{}. The latter shows strongly renormalized $xz/yz$-derived bands and a very incoherent $xy$ quasiparticle dispersion only visible around zero energy close to the X point. For \SRO{} the Hund's metal physics is responsible for the weight on the unoccupied side above $\sim\SI{0.3}{eV}$ on the X-M path. 

\begin{figure}[t]
	\centering
	\includegraphics[width = \columnwidth]{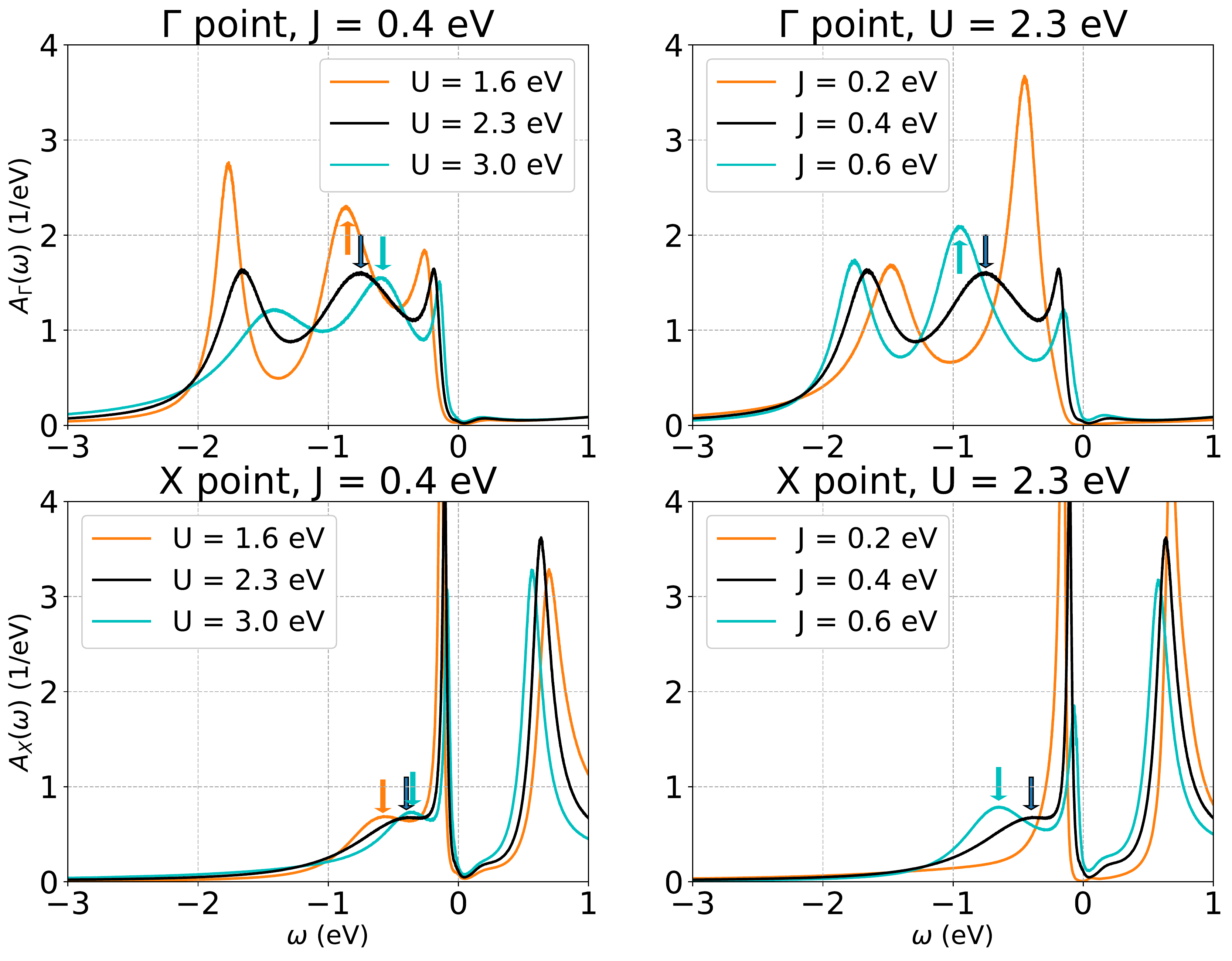}
	\caption{DMFT spectral function of \SMO{} at the $\Gamma$ point (top row) and the X point (bottom row) for different values of $U$ at fixed $J=\SI{0.4}{eV}$ (left-hand panels) and for different values of $J$ at fixed $U=\SI{2.3}{eV}$ (right-hand panels) calculated at T=\SI{232} {K}. Arrows point to the Hund's peaks.}
	\label{fig:AGw}
\end{figure}

In Fig.~\ref{fig:AGw} we examine the Hund's peak physics in more detail by focusing on the energy dependence of the spectrum at two characteristic momentum points. Concentrating first on the $\Gamma$ point, for the parameters believed to be relevant to \SMO{} and \SRO{}, a distinct three-peak structure is observed on the occupied side of the spectrum (black line). Following our discussion above, the peak closest to the chemical potential stems from the strong renormalization of the bare $xz/yz$ bands, the peak furthest from the chemical potential results from the $xy$ orbital, and the peak in the middle is a direct consequence of the Hund's metal nature of \SMO{}. Based on our calculations, the three peaks are well separated and the intensity of the Hund's metal peak is similar to the intensity of the other two peaks. 

Changing the Coulomb repulsion or Hund's coupling away from the physically expected values changes the behavior. At fixed $U$ the Hund's peak shifts away from the chemical potential with increasing $J$ (right-hand panel), while at fixed $J$ it shifts toward the chemical potential when $U$ is increased (left-hand panel). Increasing $J/U$ will favor the formation of a composite $S=3/2$ impurity spin, leading to an increased coherence energy scale for the orbital screening process, and thus the Hund's metal peak  moves to higher (negative) energies.  For too small $J/U$ outside the Hund's metal regime the three-peak structure ceases to exist, as is indeed the case for $U=\SI{2.3}{eV}$ and $J=\SI{0.2}{eV}$; see Fig.~\ref{fig:AGw} right-hand panel. We see similar behavior at the X point (Fig.~\ref{fig:AGw}, bottom panels), although the Hund's side peak is much less pronounced than at the $\Gamma$ point. For $U=\SI{2.3}{eV}$ and $J=\SI{0.4}{eV}$ (black lines), we find a small Hund's side peak at $\omega \sim \SI{-0.5}{eV}$. The peak moves closer to the chemical potential for increasing $U$ (left-hand panel), while the peak moves in the opposite direction for increasing $J$.

The calculated dependence on $U$ and $J$ excludes that the observed peak originates from the atomic multiplet structure because the multiplet splitting would evolve in the opposite way as $J$ is varied. The multiplet structure has been used to determine $J$, for example, in the Mott-insulating state of \ce{Ca2RuO4}, resulting in $J=\SI{0.4}{eV}$~\cite{sutter17}. Our work shows that the position of the Hund's peak can provide information on the interaction strength in a material which is metallic and where the atomic multiplet structure is not experimentally visible.

In addition to the single-particle quantities discussed here, we also calculate the probabilities of different local configurations of the correlated Ru/Mo site (see Supplemental Material~\cite{Supp}). We find the probability of a local high-spin configuration is almost the same in the two compounds, indicating a strong similarity of the local magnetic correlations.

In this Letter we have presented a study of the correlated electronic structure of \SMO{} in comparison with that of the well-understood material \SRO{}. The difference in electron density (2 electrons per Mo versus 4 per Ru) leads to similarities (in many respects \SMO{} is the particle-hole dual of \SRO{}) but also to pronounced differences in physics, which can be used to gain understanding of the interplay between correlated electron physics and band structure aspects. For \SMO{} the van~Hove singularity is far from the chemical potential, while for \SRO{} it is very close. A comparison of the two materials therefore provides insight into the importance of van~Hove physics in Hund's metals. Perhaps of more fundamental significance, for \SMO{} the characteristic spectral features theoretically predicted to arise in Hund's metals occur on the occupied side of the electronic spectrum and should therefore be accessible to photoemission experiments. Single-crystal \SMO{} thin films have been synthesized~\cite{radetinac2011single} and we hope these experiments may soon be feasible.
In \SMO{} no superconductivity has been found to date. Nevertheless, we believe that future studies of this material could bring key insight on the importance of Hund's physics, spin-orbit coupling, the van~Hove singularity, and other band structure details for the emergence of superconductivity in \SRO{}. Study of multiparticle physics, similar to recent works on \SRO{}~\cite{Boehnke2018,Acharya2019,Strand2019,Gingras2019}, is desirable.

~~\\
We thank A.~Georges, G.~Kotliar, and J.~Mravlje for fruitful discussions. J.K. and A.J.M. acknowledge funding by the Materials Sciences and Engineering Division, Basic Energy Sciences, Office of Science, U.S. DOE. M.B., M.G., and U.S. acknowledge support by the Deutsche Forschungsgemeinschaft (DFG, German Research Foundation) under Germany's Excellence Strategy-426 EXC-2111-390814868 and by Research Unit FOR 1807 under Project No. 207383564. M.B., M.G., and U.S. thank the Flatiron Institute for its hospitality. The Flatiron Institute is a division of the Simons Foundation.

\nocite{bulla98,haegeman11,haegeman16,paeckel2019,barthel09,caffarel94,Blaha2018,Vogt1995,Vaugier2012, hubig17, systen}

\bibliographystyle{apsrev4-1}
\bibliography{references.bib}


\clearpage
\setcounter{page}{0}
\thispagestyle{empty}
\onecolumngrid

\begin{center}
\vspace{0.1cm}
{\bfseries\large Supplemental Material for \\
``Sr$_2$MoO$_4$ and Sr$_2$RuO$_4$: Disentangling the Roles of Hund's and van~Hove Physics''}\\
\vspace{0.4cm}
Jonathan Karp,$^1$
Max Bramberger,$^{2,3}$
Martin Grundner,$^{2,3}$\\
Ulrich Schollw\"ock,$^{2,3}$
Andrew J.\ Millis,$^{4,5}$
and
Manuel Zingl$^4$
\\
\vspace{0.1cm}
{\it 
$^1$Department of Applied Physics and Applied Math,\\ Columbia University, New York, NY 10027, USA\\
$^2$Arnold Sommerfeld Center of Theoretical Physics, Department of Physics,\\ University of Munich, Theresienstrasse 37, 80333 Munich, Germany\\
$^3$Munich Center for Quantum Science and Technology (MCQST), Schellingstrasse 4, 80799 Munich, Germany\\
$^4$Center for Computational Quantum Physics, Flatiron Institute, 162 5th Avenue, New York, NY 10010, USA \\
$^5$Department of Physics, Columbia University, New York, NY 10027, USA
} \\
(Dated: \today)
\vspace{0.3cm}
\end{center}

\twocolumngrid
	
\section{Method}

We perform DFT calculations using Wien2k~\cite{Blaha2018} with the standard PBE version of the GGA functional~\cite{PBE}. For both materials we use the experimentally determined $I4/mmm$ crystal structure, with 
$a = b = \SI{3.907}{\angstrom}$ and $c = \SI{12.843}{\angstrom}$ for \SMO{}~\cite{hase2003electronic}, and $a = b = \SI{3.861}{\angstrom}$ and $c = \SI{12.722}{\angstrom}$ for \SRO{}~\cite{Vogt1995}.
The DFT calculations are converged on a $21 \times 21 \times 21 $ $k$ point grid with $RKmax=7$. We use wien2wannier~\cite{wien2wannier} and Wannier90~\cite{wannier90} to construct maximally localized Wannier functions~\cite{MLWF1, MLWF2} of t$_{2g}$ symmetry on a $10 \times 10 \times 10$ $k$ point grid and employ a frozen energy window from 
[\num{-2.0}, \num{1.5}]~\SI{}{eV} for \SMO{} and
[\num{-1.8}, \num{3.0}]~\SI{}{eV} for \SRO{}.

We use Hubbard-Kanamori on-site interactions~\cite{Kanamori1963}:
\begin{multline*}
H = U \sum_l n_{l \uparrow}n_{l \downarrow} 
+ \sum_{l<l', \sigma}[U'n_{l\sigma}n_{l'\Bar \sigma}
+ (U' - J)n_{l\sigma}n_{l'\sigma} \\
- Jc^\dagger_{l\sigma} c_{l\Bar \sigma} c^\dagger_{l' \Bar \sigma} c_{l' \sigma}]
- J \sum_{l<l'} [c^\dagger_{l\uparrow}c^\dagger_{l\downarrow}c_{l'\uparrow}c_{l'\downarrow} + H.c.]
\end{multline*}
with $l\in\{xy, xz,yz\}$ and $U' = U - 2J$. With the exception of main text Fig.~\ref{fig:AGw}, we have assumed
the same $U$ and $J$ of \SI{2.3}{eV} and \SI{0.4}{eV} for both materials. These values are commonly used for \SRO{}~\cite{Mravlje2011,Stricker2014,Kim2018,Strand2019,Zingl2019,Linden2020}. We note that cRPA estimates for the interaction parameters are about 8\% higher for \SMO{}~\cite{Vaugier2012}.

We perform DMFT calculations using the TRIQS library~\cite{TRIQS} and the TRIQS/DFTTools application~\cite{TRIQS/DFTTOOLS}, using a very dense $400 \times 400 \times 400 $ $k$ point grid. The calculations are ``one-shot" DFT+DMFT, meaning that the DFT density is kept fixed and not updated. We absorb the double counting correcting into the chemical potential, as we purely work in the low-energy subspace defined by the t$_{2g}$-like Wannier orbitals.

For calculations at a set of finite temperatures between 29 and \SI{464}{K}, we solve the impurity problem using the continuous-time quantum Monte Carlo method in the hybridization expansion (CTHYB)~\cite{TRIQS/CTHYB,Gull:2011lr}. To obtain high-quality data we use a total of $\sim 10^9$ measurements in the last iteration.
All $T=\SI{0}{K}$ calculations are carried out using a matrix product states (MPS) based impurity solver~\cite{wolf15iii, Linden2020}.
Conceptually, the MPS-based solver is equivalent to impurity solvers based on exact diagonalization~\cite{caffarel94}, and thus also allows for the inclusion of spin-orbit coupling (SOC). Calculations including SOC with CTHYB are limited to high temperatures due to a severe sign problem. The results of calculations with SOC and methodological details on the MPS-based solver are provided in the following sections.

For the calculation of quantities on the real-frequency axis, we use the inversion method~\cite{Kraberger2017} to analytically continue the self-energy. We employ the stochastic analytic continuation, following the approach by Beach~\cite{beach04}.

We quantify the strength of correlations with the renormalization factor:
\begin{equation}
Z = \left( 1 - \frac{\partial \text{Im}\Sigma(\imag \omega_n)}{\partial \omega_n}\Big|_{\omega_n \to 0}\right)^{-1}.
\end{equation}
We determine $Z$ by fitting a polynomial of 4th order to the lowest 6 points of the Matsubara self-energies and extrapolate Im$\left[\Sigma(\imag\omega_n\rightarrow 0)\right]$, 
a procedure also used in Refs.~\cite{Mravlje2011,Zingl2019}.

\section{MPS-based solver}
\label{sec:mps}

For calculations at $T=\SI{0}{K}$ (with and without SOC), we use the MPS-based impurity solver in imaginary time as introduced in Ref.~\cite{wolf15iii} and already successfully applied to \SRO{}~\cite{Linden2020}. We refer to Refs.~\cite{wolf15iii,Linden2020} for methodological details. The calculations are performed using the \textsc{SyTen} toolkit~\cite{hubig17, systen}.

For numerical purposes, and to compare to CTHYB calculations, we use a discrete grid of Matsubara frequencies at a (fictitious) inverse temperature $\beta_{\text{fict}}=\SI{200}{ eV^{-1}}$. Without the inclusion of SOC, the bath consists of three SU(2) symmetric orbitals with $L_b=8$ bath sites per spin and orbital. We use five quantum numbers: the occupation parity of each orbital, the particle number, and the spin. Occupation parity is important due to the pair hopping in the Hubbard-Kanamori Hamiltonian, such that odd and even sectors are disconnected, trapping ground state searches; auxiliary small single-particle hopping terms are numerically unreliable and deteriorate the quality of results. Ground state searches result in bond dimensions of 2048. For the time evolution we use the time-dependent-variational-principle (TDVP)~\cite{haegeman11,haegeman16,paeckel2019} up to $\tau=\SI{200}{eV}$ in steps of $\Delta \tau=\SI{0.05}{eV}$, supplemented by linear prediction~\cite{barthel09} to extrapolate to larger times. To improve numerical accuracy over Dyson's equation, we determine self-energies by using the additional correlator introduced by Bulla~\etal~\cite{bulla98}. We consider DMFT loops to be converged when the largest change in the hybridization function is below $10^{-3}$. For \SMO{} the resulting self-energy is in very good agreement with CTHYB results obtained at $T=\SI{58}{K}$; see Fig.~\ref{fig:cthyb_vs_mps}. We refer to Ref.~\cite{Linden2020} for the comparison of MPS and CTHYB results for \SRO{}.

\begin{figure}[t]
	\centering
	\includegraphics[width = \columnwidth]{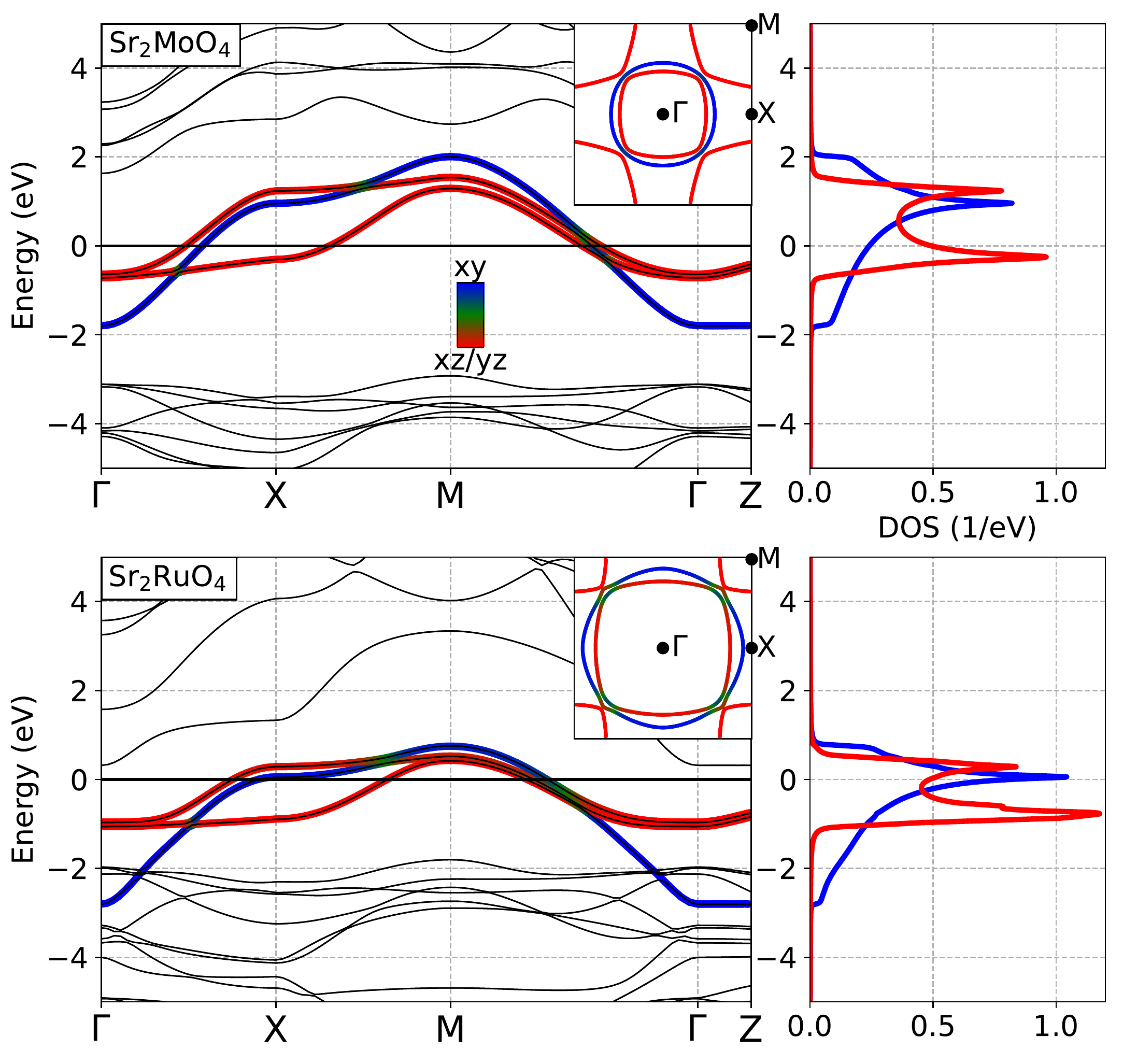}
	\caption{Left: comparison of DFT (black) and $xy$ (blue) and $xz/yz$-derived (red)  Wannier  bands for \SMO{} (top) and \SRO{} (bottom) with SOC included (see text). We add a SOC of strength $\lambda = \SI{80}{meV}$ for \SMO{} and $\lambda = \SI{100}{meV}$ for \SRO{} to the Wannier Hamiltonian. Insets: Fermi surfaces in the $k_z = 0$ plane. Right: Orbitally resolved Wannier densities of states (per spin). }
	\label{fig:soc_dft_bands}
\end{figure}

With SOC included, the determination of the Green's functions is numerically much more involved. We consider bath sizes $L_b=4$, shown in Ref.~\cite{Linden2020} to be sufficiently accurate for \SRO{} when SOC is included. The remaining quantum numbers are the particle number and the $z$ component of the total angular momentum in the $J$ basis. Ground state searches result in bond dimension of 4096, and for the imaginary time evolution two-site TDVP is used up to $\tau=\SI{100}{eV}$ in time steps of $\Delta \tau=0.05$ with subsequent linear prediction. The problem of determining a matrix-valued $(6\times 6)$-dimensional Green's function (3 bands, 2 non-degenerate spin orientations) is alleviated by rotating to the $J$ basis, reducing the problem to two $1\times1$ blocks and two $2\times 2$ blocks~\cite{Linden2020}. 

\begin{figure}[t]
	\centering
	\includegraphics[width = 1.0\columnwidth]{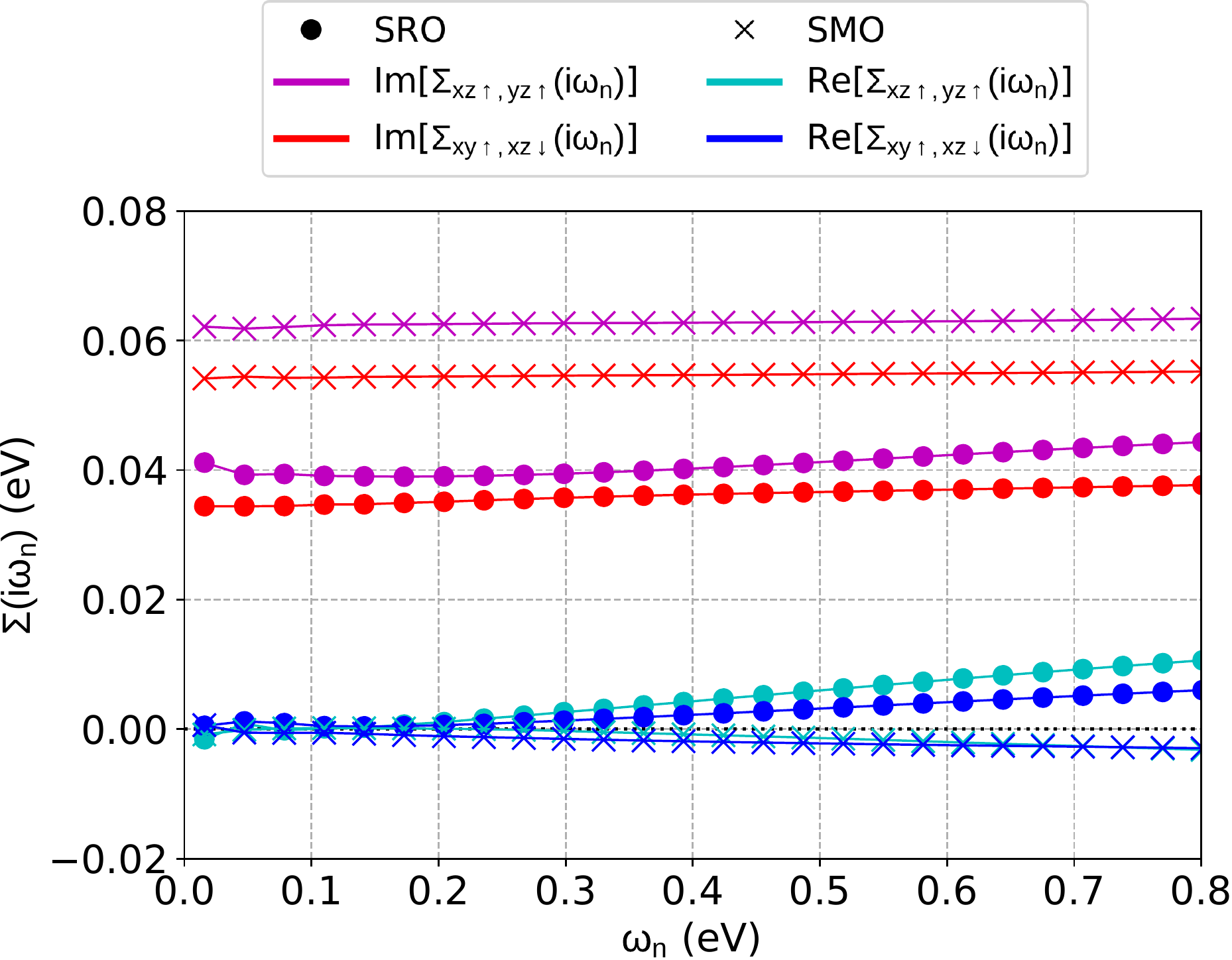}
	\caption{Selected off-diagonal elements of the self-energy for \SRO{} reproduced from Ref.~\cite{Linden2020} (circles) and \SMO{} (crosses) calculated with the MPS-based impurity solver using a bath size of $L_b = 4$ per orbital and spin state. Interaction parameters are $U = \SI{2.3}{eV}$ and $J = \SI{0.4}{eV}$.}
	\label{fig:soc_sigma_off}
\end{figure}

\section{Spin-orbit Coupling (SOC)}
\label{sec:soc}

To study the effect of SOC, we add a local t$_{2g}$-only spin-orbit term, see e.g. Refs.~\cite{Zhang2016,Linden2020}, to the Wannier Hamiltonian. We determine the SOC strength $\lambda$ by matching the resulting eigenenergies to the Kohn-Sham bands obtained from DFT calculations with SOC included; see Fig.~\ref{fig:soc_dft_bands}. Using SOC strengths of $\lambda^{\text{SRO}}_{\text{DFT}}=\SI{100}{meV}$ for \SRO{} and $\lambda^{\text{SMO}}_{\text{DFT}}=\SI{80}{meV}$ for \SMO{} results in a nearly perfect agreement with DFT. For \SRO{}, the SOC substantially reshapes the Fermi surface, cf. Fig.~\ref{fig:dft_bands} of the main text and Fig.~\ref{fig:soc_dft_bands} here, which has been also shown, for example, in Refs.~\cite{Haverkort2008,Zhang2016, Kim2018, Tamai2018}.
In contrast to \SRO{}, the SOC does not distort the shape of the Fermi surface for \SMO{}.

Additionally, the SOC leads to a mixed $xy$-$xz/yz$ orbital character of degenerate and nearly degenerate states.
On the Fermi surface, the mixed orbital character is especially important along the diagonal direction $\Gamma$-X; see inset of Fig.~\ref{fig:soc_dft_bands}. For \SRO{}, this results in states on the Fermi surface with 50:50 mixing of $xy$ and $xz/yz$ characters, while for \SMO{} the mixing of the orbital character is less pronounced. The reason for this lies in the fact that the SOC is only effective in hybridizing orbitals when bands cross or are close in energy. This occurs at different $k$ points and over a greater range of $k$ space for \SRO{}; cf. the $\Gamma$-M path shown in Fig.~\ref{fig:soc_dft_bands} and Fig.~\ref{fig:dft_bands} of the main text.

For \SRO{} it has been predicted theoretically~\cite{liu_prl_2008,Zhang2016,Kim2018} and confirmed experimentally~\cite{Tamai2018} that electronic correlations lead to an enhancement of the effective SOC. To be precise, the off-diagonal elements of the self-energy have the same structure as the SOC term and are found to be almost frequency independent. Hence, the physics can be described -- to a good approximation -- by an effectively enhanced SOC term ($\lambda_{\text{eff}} = \lambda_{\text{DFT}} + 2\Sigma_{\text{off.}}(\omega = 0)$, following the definition of $\lambda$ in Ref.~\cite{Linden2020}) and a purely diagonal self-energy. For \SRO{}, the MPS-based impurity solver yields a correlation-enhanced effective SOC of $\lambda^{\text{SRO}}_{z}=\SI{192}{meV}$ and $\lambda^{\text{SRO}}_{xy}=\SI{179}{meV}$~\cite{Linden2020}, which is nearly 2 times larger than $\lambda_{\text{DFT}}$. For \SMO{}, we find similar correlation-enhanced SOC of $\lambda^{\text{SMO}}_{z}=\SI{204}{meV}$ and $\lambda^{\text{SMO}}_{xy}=\SI{182}{meV}$, see Fig.~\ref{fig:soc_sigma_off}, while the bare SOC is \SI{20}{meV} smaller than for \SRO{}. Thus, at the same interaction values of $U$ and $J$, electronic correlations enhance the SOC by a factor of about 2.5, which is even more than what is found for \SRO{}. The enhancement of the SOC is affected both by the intrinsic local interaction strength and by the degree to which the bands are entangled~\cite{liu_prl_2008}. The former is the same for both compounds in our calculations, while the latter, we believe, is what leads to the difference found in the off-diagonal elements of the self-energy.

\begin{figure}[t]
	\centering
	\includegraphics[width = \columnwidth]{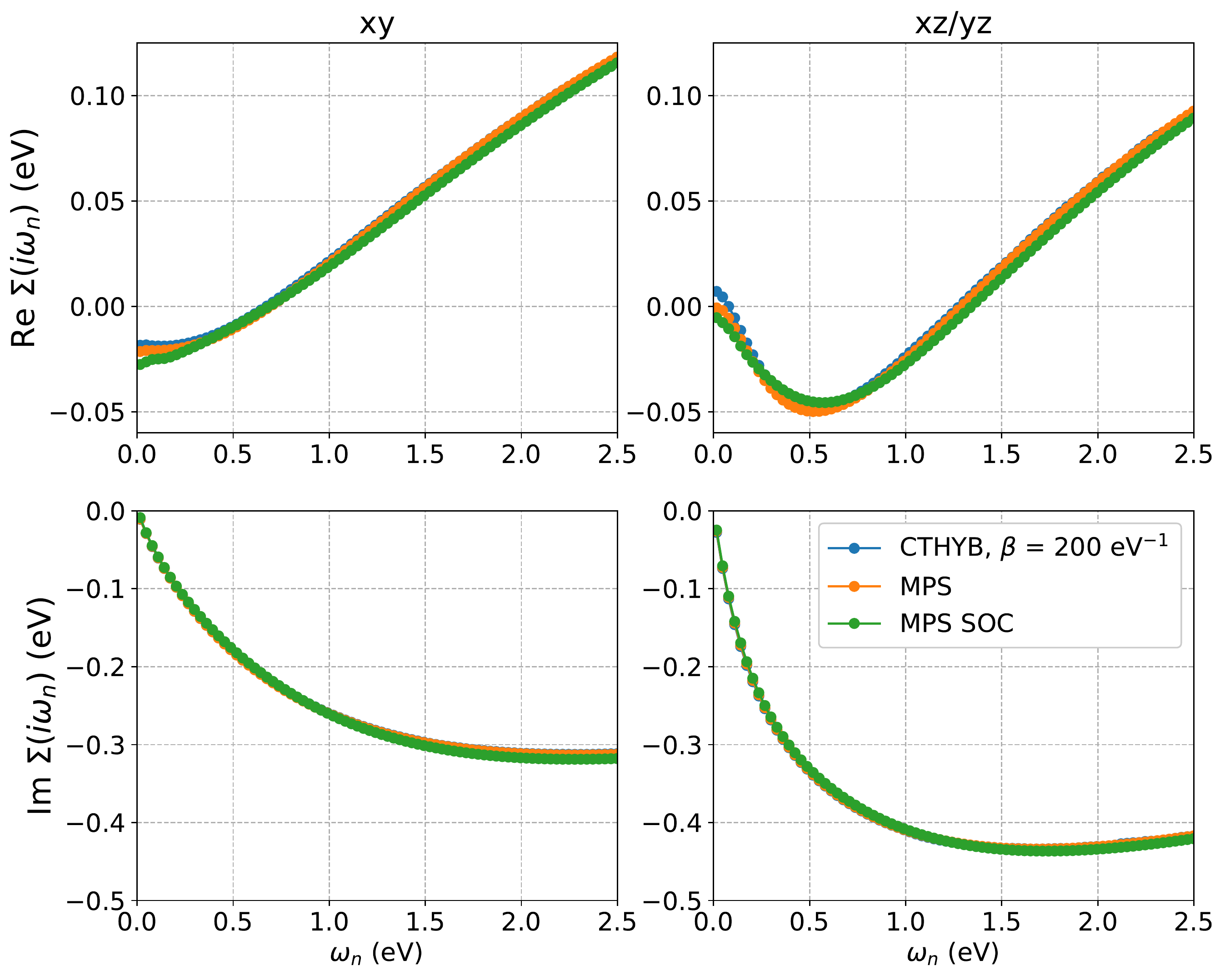}
	\caption{Comparison of Matsubara self-energies for \SMO{} obtained with the MPS-based impurity solver at $T=0$ with (orange) and without (green) SOC compared to CTHYB results obtained without SOC at $T=\SI{58}{K}$. Interaction parameters are $U = \SI{2.3}{eV}$ and $J = \SI{0.4}{eV}$. We subtracted the chemical potential from the real parts.}
	\label{fig:cthyb_vs_mps}
\end{figure}

On the other hand, the diagonal elements are nearly unchanged by the inclusion of SOC, as shown in Fig.~\ref{fig:cthyb_vs_mps} for \SMO{} and in Ref.~\cite{Linden2020} for \SRO{}. This implies that the mass enhancements of both materials are not influenced by the SOC.

\section{Imaginary part of self-energy}

\begin{figure}[t]
	\centering
	\includegraphics[width = \columnwidth]{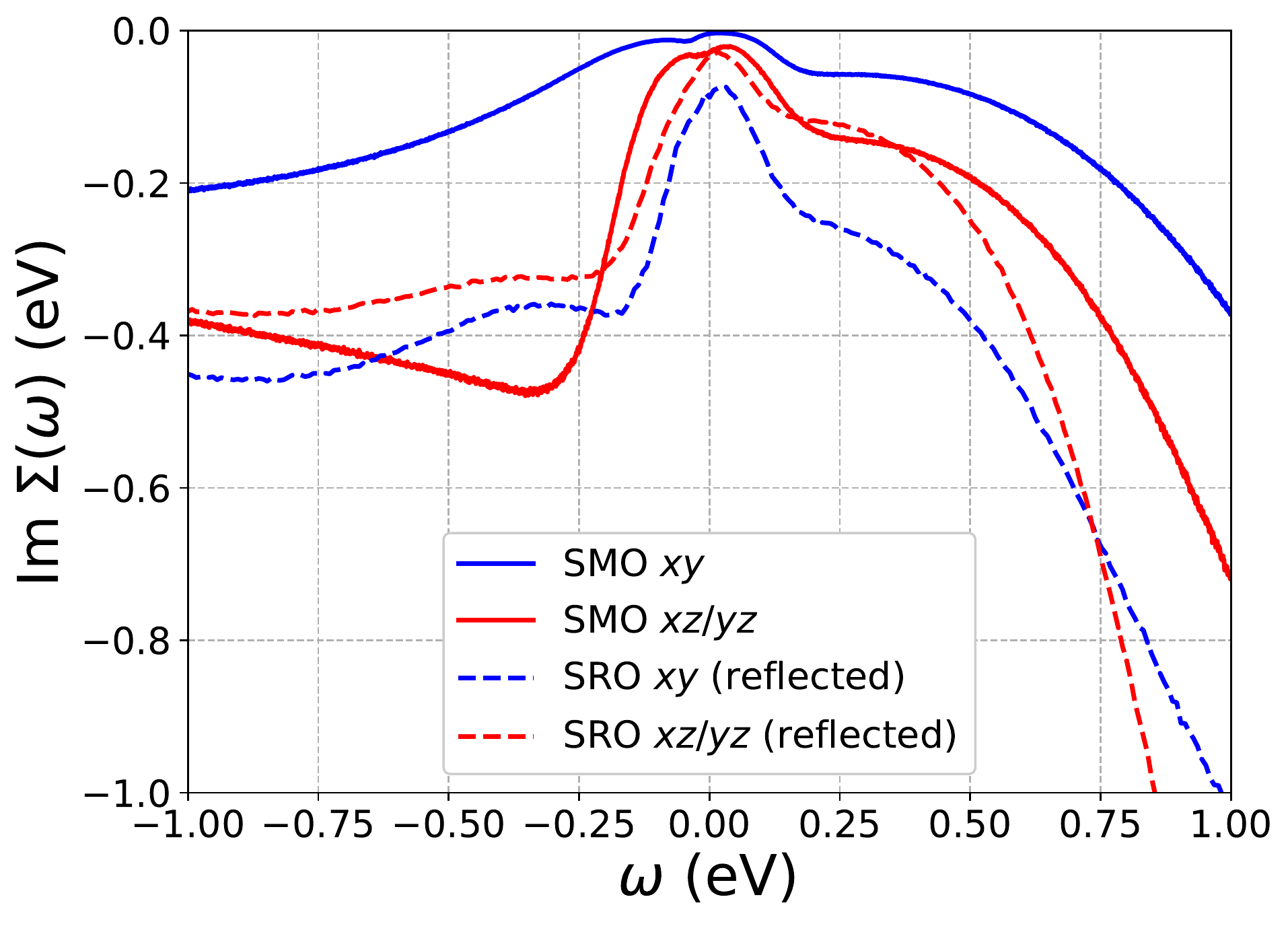}
	\caption{Imaginary part of the real-frequency self-energy for both materials at $T = \SI{232}{K}$ using CTHYB as impurity solver~\cite{TRIQS/CTHYB} with $U = \SI{2.3}{eV}$ and $J = \SI{0.4}{eV}$ and analytic continuation~\cite{beach04} to the real-frequency axis. For \SRO{} (dashed lines) we show the reflection of the self-energies through $\omega=0$; i.e., $\Sigma(-\omega)$. The real parts are shown in the inset of main text Fig.~\ref{fig:sigma}.}
	\label{fig:Im_Sigma}
\end{figure}

Fig.~\ref{fig:Im_Sigma} shows the imaginary part of the real-frequency self-energy (corresponding to the real part shown in Fig.~\ref{fig:sigma} of the main text) at $T=\SI{232}{K}$. For \SRO{} what is shown is the reflection of the self-energy through $\omega=0$; i.e., $\Sigma(-\omega)$. Like the real part, the imaginary part of the \SRO{} $xz/yz$-orbital self-energy has a similar structure to that of \SMO{}. We also note that the \SMO{} $xy$ orbital self-energy has a smaller imaginary part at $\omega = 0$ than the $xz/yz$ orbitals, opposite to what we find for \SRO{}. In the latter, the substantial electron-electron scattering in the $xy$ orbital and the presence of the van~Hove singularity at the chemical potential leads to a broadening of the $xy$-derived many-body Fermi surface sheet at finite temperature; see insets of main text Fig.~\ref{fig:correlated_structure}. For \SMO{}, where the van~Hove singularity is not in proximity to the chemical potential, the $xy$ orbital is much more coherent with a smaller imaginary part of the self-energy at $\omega = 0$, resulting in a very sharp many-body Fermi surface. 

\section{Quasiparticle Dispersions}

\begin{figure}[b]
	\centering
	\includegraphics[width = \textwidth]{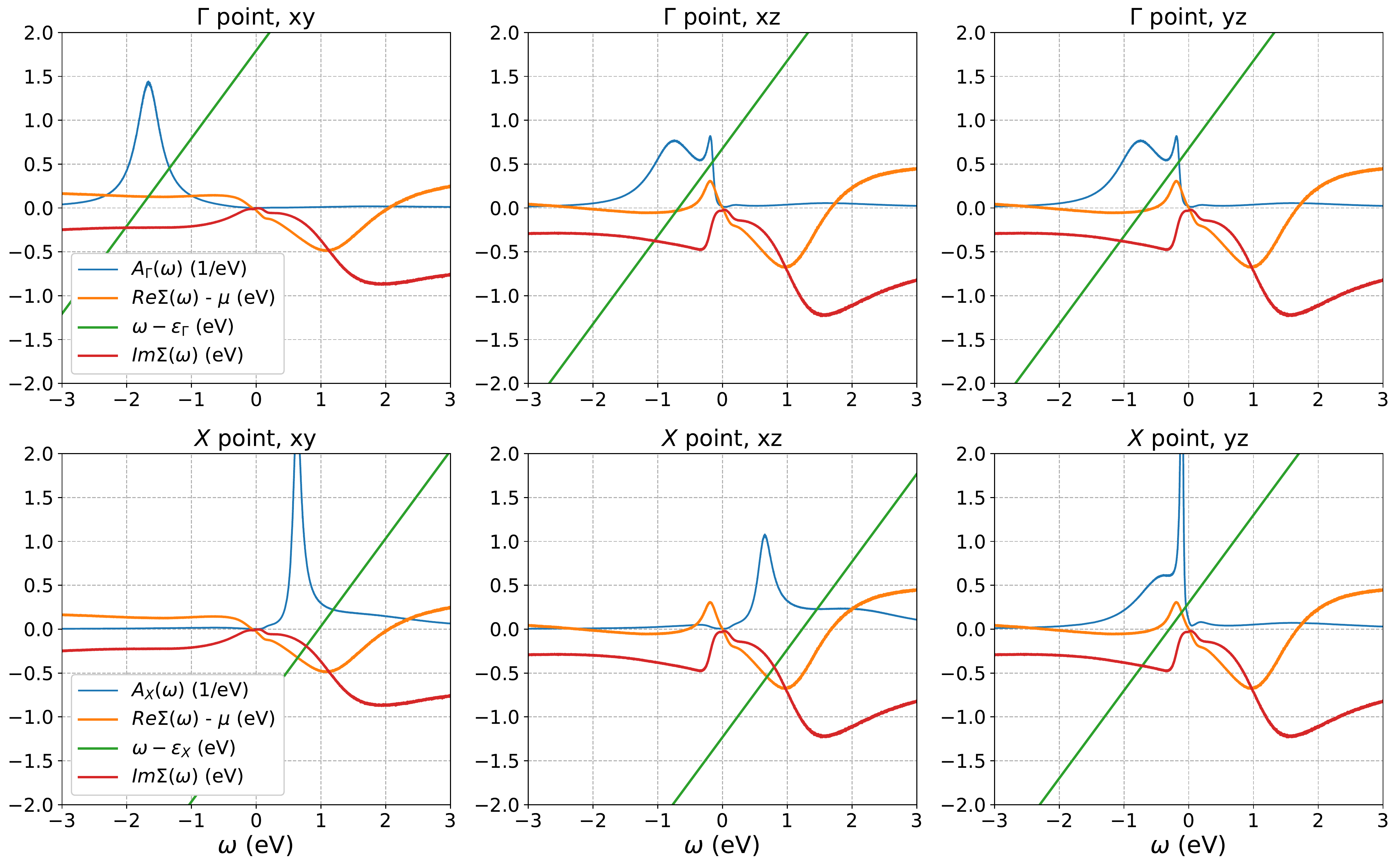}
	\caption{Relationship of the inverted slope in the real part of the self-energy to the peaks in the spectral functions for \SMO{} at $T=\SI{232}{K}$. The orbitally resolved spectral functions (blue lines) are shown together with the real parts (orange lines) and imaginary parts (red lines) of the self-energy and the linear function $\omega - \epsilon_l(k)$ (see text for further details). The top row shows the $\Gamma$ point and the bottom row the X point, and the columns correspond to the different orbitals.}
	\label{fig:qp_dispersions}
\end{figure}

To better understand how the peculiar structure in the real part of the self-energies leads to an additional peak in the spectral function, we look at the quasiparticle dispersions, given by 
$\det\left[(\omega +\mu)\delta_{l,l'} - H_{l,l'}(k) - \text{Re}\left[\Sigma_{l}(\omega)\delta_{l,l'}\right] \right]=0$. Note that without SOC the self-energy in the orbital basis is diagonal. We focus only on the $\Gamma$ and X points and take into account that for the studied materials at those point this equation simplifies to $\omega - \epsilon_l(k) = \text{Re}\left[\Sigma_{l}(\omega)\right] - \mu$ due to $H_{l,l'}(\Gamma,\text{X}) = \epsilon_l(\Gamma,\text{X}) \delta_{l,l'}$. The left-hand side, $\omega - \epsilon_l(k)$, of the equation is shown as straight lines in Fig.~\ref{fig:qp_dispersions} and the intersections with $\text{Re}\left[\Sigma_{l}(\omega)\right] - \mu$ give the solutions of the quasiparticle equation, which yield peaks in the corresponding spectral functions. There are also peaks that can emerge when the quasiparticle equation is almost satisfied, given that the imaginary part of the self-energy introduces a large enough broadening of the peaks. For example, at the $\Gamma$ point this leads to a peak at about \SI{-0.2}{eV} in the $xz$ and $yz$ orbitals. We emphasize that this peak would prevail even without the inversion of slope in the self-energy. On the contrary, the second peak at about \SI{-0.8}{eV} is a direct consequence of the inverted slope (together with the fact that it stays flat at further negative frequencies), and hence this peak is a signature of the Hund's metal physics governing \SMO{}.

\section{Ru/Mo Configurations}
\begin{figure}
	\centering
	\includegraphics[width = \textwidth]{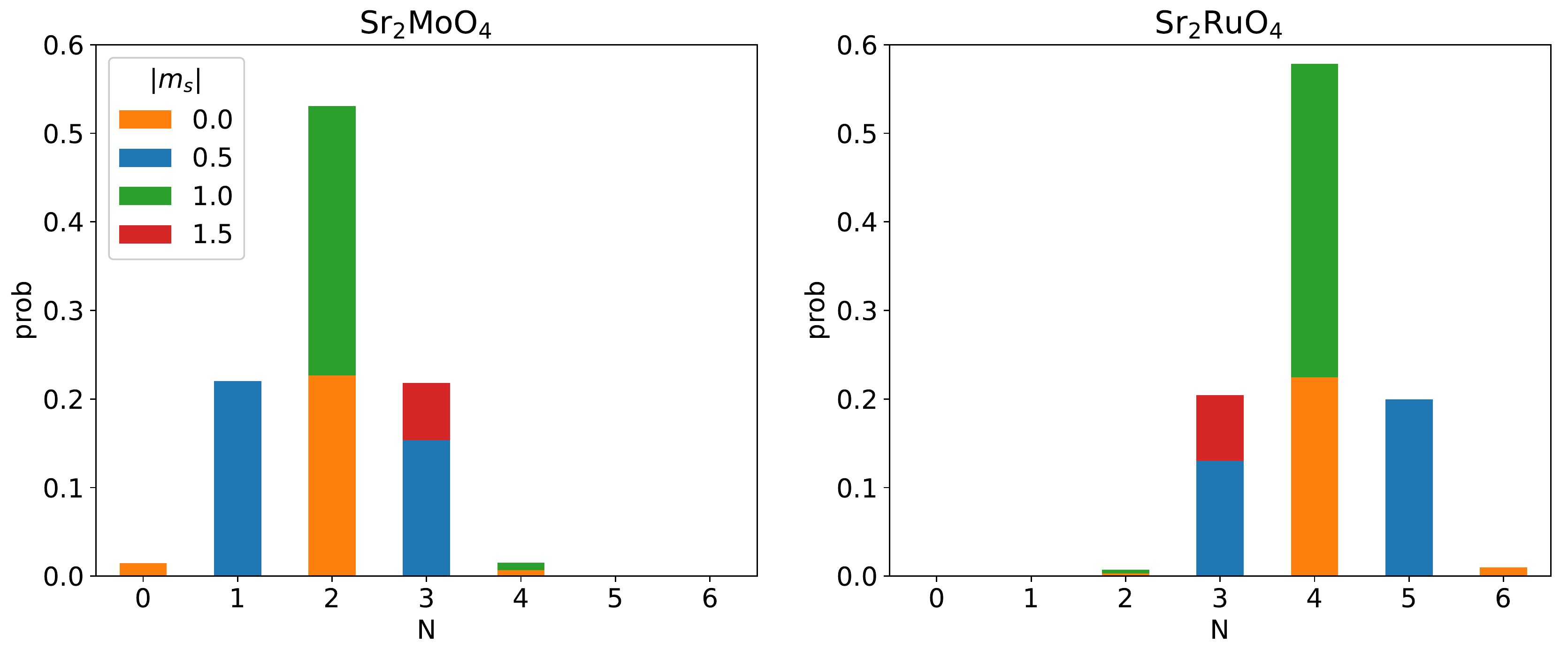}
	\caption{Occupation probability histograms of electronic configurations on the impurity calculated with the CTHYB solver~\cite{TRIQS/CTHYB} at $T = \SI{232}{K}$ using $U = \SI{2.3}{eV}$ and $J = \SI{0.4}{eV}$. The histograms are resolved into particle number $N$ and spin $|m_s|$.}
	\label{fig:hist}
\end{figure}

By measuring the frequency of occupation of each Fock state in CTHYB, we can construct a density matrix in Fock space. From the density matrix, we obtain the probability distributions of the different many body configurations of the correlated ions, shown in Fig.~\ref{fig:hist}. Overall, the histograms of \SRO{} and \SMO{} are approximately particle-hole duals, which underlines the similarity of these two compounds. For both materials, we find that the dominant configuration is 2 electrons (\SMO{}) or 2 holes (\SRO{}), and these carrier configurations are dominantly but not exclusively in the high-spin state, with the fraction being slightly higher for \SRO{}. This indicates that the local spin physics of the two materials is similar. The admixture of low-spin states likely reflects the competition between Hund's physics and crystal field splitting. For $N = 3$, we find a large amount of spin $1/2$ but still a substantial amount of spin $3/2$, corresponding to the Hund's peaks.

\onecolumngrid

\clearpage

\end{document}